\documentclass[12pt]{article}
\usepackage{amsfonts}
\usepackage{graphicx}
\usepackage{epsfig}
\usepackage{amsmath}
\usepackage{amssymb}
\usepackage{caption2}
\begin{document}
\begin{titlepage}
\begin{center}
{\Large {\bf 
{Vector boson production in association with KK modes of the 
ADD model to NLO in QCD at LHC
}}} 
\\[1cm]

M.\ C.\ Kumar$^{a}, $\footnote{mckumar@hri.res.in} 
\hspace{.5cm} 
Prakash Mathews$^{b}, $\footnote{prakash.mathews@saha.ac.in}
\hspace{.5cm} 
V.\ Ravindran$^{a}, $\footnote{ravindra@hri.res.in}
\hspace{.5cm} 
Satyajit Seth$^{b}, $\footnote{satyajit.seth@saha.ac.in}
\\[1cm]

${}^a$ Regional Centre for Accelerator-based Particle Physics\\ 
Harish-Chandra Research Institute, Chhatnag Road, Jhunsi,\\
 Allahabad 211 019, India
\\[.7cm]

${}^b$ Saha Institute of Nuclear Physics, 1/AF Bidhan Nagar, Kolkata 700 064, 
India
\end{center}
\vspace{1cm}

\begin{abstract}
\noindent
Next-to-leading order QCD corrections to the associated production 
of vector boson ($Z$/$W^\pm$) with the the Kaluza-Klein modes of the 
graviton in large extra dimensional model at the LHC, are presented.  
We have obtained various kinematic distributions using a Monte Carlo 
code which is based on the two cut off phase space slicing method that 
handles soft and collinear singularities appearing at NLO level.
We estimate the impact of the QCD corrections on various observables 
and find that they are significant. We also show the reduction in 
factorization scale uncertainty when QCD corrections are included.
\end{abstract}


\vspace{1cm}
\end{titlepage}

\section{Introduction}

With the on set of the Large Hadron Collider (LHC) era, unique opportunity 
to probe the realm of new physics in the TeV scale has begun.  Models with
extra spatial dimensions and TeV scale gravity, are proposed to address the 
large hierarchy between the electroweak and Planck scale and are expected
to provide a plethora of new and interesting signals.  

The extra dimension model proposed by Arkani-Hamed, Dimopoulos 
and Dvali (ADD) \cite{ADD}, was the first extra dimension model 
in which the compactified dimensions could be of macroscopic 
size and consistent with present experiments.  A viable mechanism 
to hide the extra spatial dimension, is to introduce a 3-brane 
with negligible tension and localise the Standard Model (SM) 
particles on it.  
Only gravity is allowed to propagate in the full $4+\delta$ dimensional 
space time.  As a consequence of these assumptions, it follows 
from Gauss Law that the effective Planck scale $M_P$ in 4-dimensions 
is related to the $4+\delta$ dimensional fundamental scale $M_S$ through 
the volume of the compactified extra dimensions \cite{ADD}.  The 
extra dimensions are compactified on a torus of common circumference 
$R_S$.  The compactified extra dimensions are flat, of equal size and 
could be large.  The large volume of the compactified extra spatial 
dimensions would account for the dilution of gravity in 4-dimensions 
and hence the hierarchy.  Current experimental limits on deviation
from inverse square law \cite{expt}, constraint the number of possible 
extra spatial dimensions $\delta \ge 2$.  The space time is factorisable 
and the 4-dimensional spectrum consists of the SM confined to 
4-dimensions and a tower of Kaluza-Klein (KK) modes, of the 
graviton propagating in the full $4+\delta$ dimensional space time.

The interaction of the KK modes 
$G_{\mu\nu}^{(\vec n)}$ 
with the SM fields localised on the 3-brane is described by an 
effective theory given by \cite{GRW,HLZ}
\begin{eqnarray}
{\cal L}_{int} &=& - \frac{1}{\overline M_P} \sum_{\vec n=0}^\infty 
T^{\mu\nu} (x) ~G_{\mu\nu}^{(\vec n)} (x) ~,
\end{eqnarray}
where $T^{\mu\nu}$ is the energy-momentum tensor of the SM fields on the 
3-brane and 
$\overline{M}_P=M_P/
\sqrt{8 \pi}$ is the reduced Planck mass in 4-dimensions.  
The relation between the 4-dimensional coupling, the volume of the extra
dimensions and the fundamental scale $M_D$ in $4+\delta$-dimensions
\begin{eqnarray}
\overline M_P^2 &=& R_D^\delta M_D^{\delta +2} ~.
\end{eqnarray}
The size of the extra dimension $R_S$ is related to the radius $R_D$, 
$R_S = 2 \pi R_D$.   
The fundamental scales in $4+\delta$ dimensions, as defined in \cite{GRW} 
$M_D$ is related to $M_S$ \cite{HLZ} as: $ 8 \pi M_D^{\delta+2} = 
M_S^{\delta+2} S_{\delta -1}$, where 
$S_{\delta -1}=2 \pi^{\delta/2}/\Gamma(\delta/2)$ is the surface
area of a unit sphere in $\delta$ dimensions.

The zero mode corresponds to the 
usual 4-dimensional massless graviton and higher massive KK modes are 
labeled by $\vec n =(n_1,n_2, \cdots , n_\delta)$.  The masses of the 
individual KK modes are $m_{\vec n}^2= {\vec n}^2/R_D^2$ and 
the mass gap between adjacent KK modes is $\Delta m=R_D^{-1}$. For 
not too large $\delta$ the discrete mass spectrum could be replaced by 
a continuum, with the density of KK states 
\begin{eqnarray}
\rho (m_{\vec n}) = \frac{1}{2} S_{\delta-1} R_D^\delta m_{\vec n}^{\delta-2} 
~. 
\label{d_state}
\end{eqnarray}
For an inclusive 
cross section at the collider, we have to sum over all accessible KK modes 
and hence cross section for the production of an individual KK mode of mass 
$m_{\vec n}$ has to be convoluted with the density of states 
$\rho (m_{\vec n})$.  The discrete sum $\sum_{\vec n}$ can be replaced by 
$\int d m^2 \rho (m_{\vec n})$, and hence the inclusive cross section
for the tower of KK modes is 
\begin{eqnarray}
{d \sigma}&=&  
S_{\delta-1} {\overline M_P^2 \over M_D^{2+\delta}}
\int dm ~m^{\delta-1} ~ {d \sigma^{(m)}_D} ~,
\label{d_state2}
\end{eqnarray}
where $d\sigma_D^{(m)}$ is the cross section to produce a single KK mode.
The cross section for an individual KK mode is suppressed by the 
coupling factor 
$(2 \overline M_P^2)^{-1}$, 
the high multiplicity of 
accessible KK modes at the collider would compensate, leading to the 
exciting possibility of observing low scale quantum gravity effects at 
the LHC.  The additional $1/2$ factor 
is due 
to the definition of the sum of polarisation of the KK modes \cite{GRW}.  
The $\sum_{\vec n}$ is 
kinematically 
constrained to those KK modes which satisfy $m_{\vec n}=|\vec n|/R_D 
< \sqrt{s}$, where $\sqrt{s}$ is the partonic center of mass energy or as 
the case may be the available energy to produce the KK modes.

Viable signatures of the ADD scenario at the LHC are possible by the
exchange of virtual KK modes between the SM particles, leading to an
enhanced cross section or by the emission of real KK modes from the
SM particles, leading to a missing energy signal.  Various such 
processes have been extensively studied in this model, most of which 
have been considered only up to leading order (LO) in QCD \cite{GRW,HLZ,
Peskin,us,Wu}.  These LO 
approximations
at the hadron colliders suffer from large factorisation and 
renormalisation scale dependence which for some processes could be
as large as a factor of two.  These issues go beyond normalisation
of a cross section by a $K$-factor as the shapes of distributions
may not be modeled correctly and in addition the LO cross sections
are strongly dependent on the factorisation scale.
It is hence essential to evaluate the 
next-to-leading order (NLO) corrections to the process of interest to 
provide quantitatively reliable predictions.  NLO QCD corrections 
to extra dimension models have been studied for dilepton \cite{us2}, 
boson pair \cite{us3,us4} productions, and real graviton production 
processes such as graviton plus jet \cite{KKLZ} and graviton plus photon 
\cite{GLGW}.  Searches at the Tevatron using the single photon or 
jet with missing transverse energy have been used to put bounds on
extra dimensional scale $M_D$ for different number of extra dimensions
\cite{exptX1,exptX2}.  The same signal has been simulated for the 
LHC at the ATLAS detector \cite{Vacavant}, discovery limit and the 
methods to determination of the parameters of the extra dimensional
models are discussed.

In this paper we consider the graviton production in association with
a vector boson at the hadron colliders at NLO in QCD.  Z-boson process 
to LO had been considered at LEP \cite{Kingman} and 
simulation studies for the $Z+G_{KK}$ modes production at the LHC 
was studied to LO \cite{Ask} as a complement to the more conventional 
channels.  

\section{Vector bosons in association with KK modes}

The associated production of Z-boson with the KK modes of the ADD 
model leads to missing energy signals at the hadron colliders. 
We begin by discussing the neutral weak gauge boson ($Z$)
production in association with the KK modes of the ADD large extra 
dimensional model to NLO in QCD and would consider the charged weak 
gauge bosons ($W^\pm$) towards the end.

At the hadron collider, the associated production $P ~P \to Z ~G_{KK} 
~X$ at LO proceeds via the quark, anti-quark annihilation process $q 
~\bar q \to Z ~G_{KK}$.  There are four diagrams that would contribute 
to this process, which corresponds to the KK modes of the graviton being
emitted of a fermion leg, Z-boson or the $q ~\bar q ~Z$ vertex.  The 
Feynman rules to evaluate the matrix elements are given in \cite{GRW, 
HLZ} and for the vector boson, unitary gauge ($\xi \to \infty$) is used.  
Summation of the polarisation tensor of the KK modes is given in 
\cite{GRW}.
It can be seen that the terms proportional 
to the inverse powers of KK mode mass $m^2$ vanish on expressing the 
matrix element square in terms of independent variable and is an useful 
check.  

The NLO calculation presented here uses both analytical and Monte Carlo 
integration methods and hence is flexible to incorporate the experimental 
cuts and can generate the various distributions unlike a fully analytical 
computation.  Our code is based on the method of two cutoff phase space 
slicing \cite{Harris} to deal with various singularities appearing in the NLO 
computation of the real diagrams and to implement the numerical 
integrations over phase space.  The analytical results are evaluated 
using the algebraic manipulation program FORM \cite{FORM}.  The real and 
virtual corrections have been evaluated in the massless quark limit.  
We use dimensional regularisation with space time dimensions 
$d=4+\epsilon$.  To deal with $\gamma_5$ in $d$-dimensions, we use the 
completely anti-commuting $\gamma_5$ prescription \cite{CFH}.

The order ${\cal O} (\alpha_s)$ corrections to the associated production
of the Z-boson and KK modes of the graviton come from the following 
$2 \to 3$ real diagrams 
(a) $q ~\bar q \to Z ~G_{KK} ~g$, (b) $q (\bar q) ~g \to q (\bar q) 
~Z ~G_{KK}$.  
There are 14 diagrams that contribute to the quark antiquark annihilation 
process and can be classified into diagrams where the KK modes couples 
to the 
(i) fermion legs, (ii) Z-boson leg, (iii) gluon leg, (iv) $q ~\bar q ~Z$ 
vertex and (v) $q ~\bar q ~g$ vertex.  The unitary gauge is used for the
Z boson propagator and Feynman gauge for the gluon propagator. 
Note that the gauge parameters influence the coupling of KK modes to the
SM fields \cite{HLZ}.
For the sum over 
the polarisation vectors we have retained only the physical degrees of 
freedom.  

The real diagrams involving gluons and massless quarks would be singular 
in soft and collinear regions of the 3-body phase space integration.
Two small slicing parameters $\delta_s$ and $\delta_c$ 
are introduced to isolate regions of phase space that are sensitive to 
soft and collinear singularities.  Rest of the region is finite and
can be evaluated in 4-dimensions.  Phase space integrations in the 
mutually exclusive soft and collinear regions are performed not on the 
full matrix element but in the leading pole approximation of soft
and collinear regions in $4+\epsilon$ dimensions.  The soft and collinear
poles now appear as poles in $\epsilon$ and in addition the soft part 
would depend logarithmically on the soft cut-off $\delta_s$ while the 
collinear part would depend on the both $\delta_s$ and $\delta_c$.  All 
positive powers of the small cut-off parameters are set to be zero.  The 
phase space degrees of freedom that remain, correspond to a 2-body 
process and can now be combined with the virtual diagram.

The virtual corrections to the annihilation process to order ${\cal O} 
(\alpha_s)$ can be obtained by considering the gluonic virtual 
corrections to the vertex and wave function renormalisation for the 
process $q \bar q 
\to Z$ and then attaching the KK modes to all possible legs and vertex 
as allowed by the Feynman rules \cite{GRW,HLZ}.  This would generate 
27, one loop 
$2 \to 2$ diagrams which on multiplication with the $q ~\bar q ~\to 
~Z ~G_{KK}$ at LO would give all the virtual contributions for the 
annihilation process.  Performing the loop integrals in $4+\epsilon$
dimensions would lead to poles in $\epsilon$ in the soft and collinear
regions and combining this with the real diagrams would lead to the
cancellation of the soft singularities.  The remaining collinear singular 
terms which appear as poles in $\epsilon$ are systematically removed by 
collinear counter terms in the $\overline{MS}$ factorisation scheme, at an
arbitrary factorisation scale $\mu_F$.  The resultant expression is now
finite but depends on the cut-off parameters $\delta_s$ and $\delta_c$
logarithmically.  Combining this $2 \to 2$ part with the finite $2 \to 3$
part and performing the phase space integration for the various 
distributions, it is expected that the results be independent of the 
choice of the slicing parameters $\delta_s$ and $\delta_c$.

The $q (\bar q) ~g ~\to ~q (\bar q) ~Z ~G_{KK}$ process begins
at ${\cal O} (\alpha_s)$ so does not get any virtual corrections to this
order and can be obtained from the $2 \to 3$ annihilation diagrams by using 
the crossing symmetry.  
The analytical results of LO $2 \rightarrow 2$, finite part of
NLO virtual corrections and full $2 \rightarrow 3$ matrix elements
that go into our numerical code will be presented in the longer version
\cite{future}.

\section{Numerical results}

In this section we present the numerical results for the associated
production of the $Z$-boson with the KK modes at the LHC ($\sqrt{S} = 14$ TeV)
to NLO in QCD.   The mass of the $Z$-boson and the weak mixing angle
are taken to be $m_Z = 91.1876$ and $\hbox{sin}^2 \theta_w = 0.2312$ 
respectively.  The fine structure constant is taken to be $\alpha = 1/128$.
Through out our computation we have used CTEQ6 L/M parton density 
sets \cite{cteq}
and $n_f = 5$ light quark flavors.  The two loop running strong coupling 
constant 
in the $\overline{MS}$ scheme has been
used with the corresponding $\Lambda_{QCD} = 0.226$ GeV.  
Unless mentioned other wise, both the renormalization and the factorization 
scales are set to $\mu_R = \mu_F = p_T^Z$, the transverse momentum of the 
$Z$-boson. Further, the following cuts have been implemented in our numerical
code:
\begin{eqnarray}
p_T^Z, ~ p_T^{miss}  >  p_T^{min} \quad \hbox{ and } |y^Z|  <  2.5 
\end{eqnarray}
where $y^Z$ is the rapidity of the $Z$-boson, $p_T^{min} = 400$ GeV and
the missing transverse momentum $p_T^{miss}$ is given by
\begin{eqnarray}
p_T^{miss} = p_T^Z ~(p_T^G) 
\quad \quad 
\hbox{for $p_T^{jet}  <  20$ ( $> 20$) {\rm GeV}}
\label{ptmiss}
\end{eqnarray}
At LO, the missing transverse momentum $p_T^{miss}$ is the same as $p_T^Z$.  
The additional jet at NLO can be soft or hard; For the soft jet the 
$p_T^{miss}$ is the same as $p_T^Z$ while for hard jet $p_T^{miss}$ is 
$p_T^G$ of the KK modes (Eq.\ (\ref{ptmiss})).  In addition 
to the above, we have put a cut on the pseudo rapidity of the jet 
$|\eta^{jet}| < 2.5$ for $p_T^{jet} > 20$ GeV.  
\begin{figure}[htb]
\centerline{
\epsfig{file=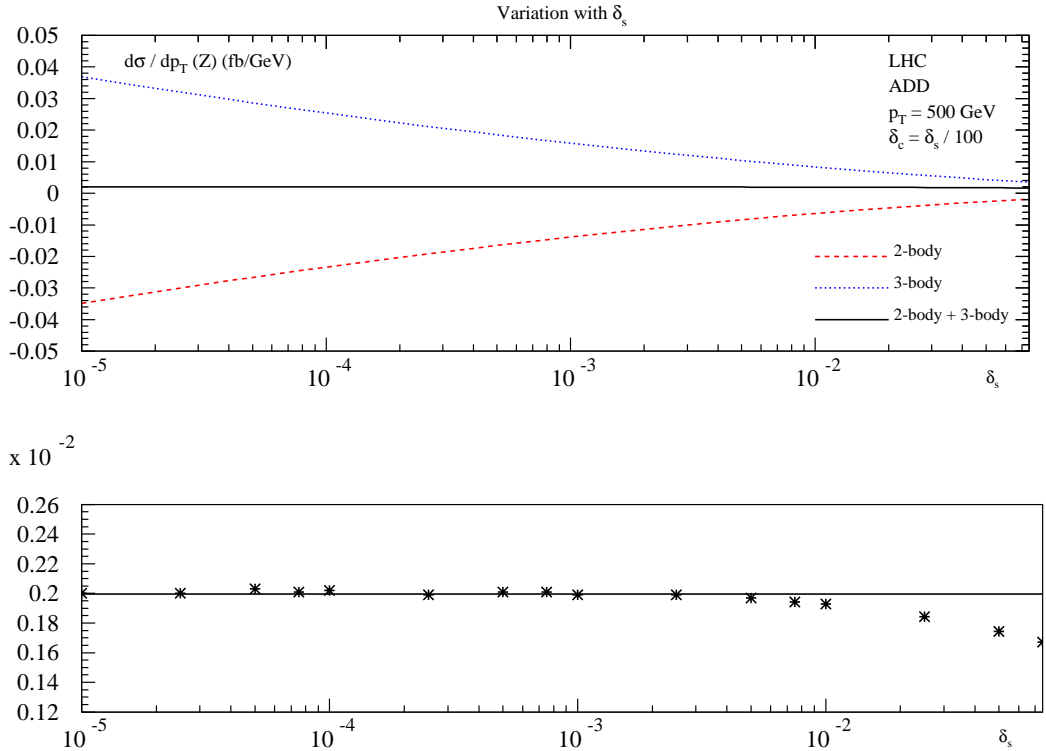,width=15cm,height=12cm,angle=0}}
\caption{Dependence of the order $\alpha_s$ contribution to the transverse 
momentum distribution of the $Z$-boson at the LHC, on the slicing parameter 
$\delta_s$ (top) with $\delta_c = \delta_s /100$ and for $M_D = 3$ TeV and 
$\delta = 4$.  Below the variation the sum of $2$-body and $3$-body
contributions is contrasted against the value at $\delta_s = 10^{-3}$.}
\label{deltas}
\end{figure}

We check the stability of our results 
against the variation of the slicing parameters $\delta_s$ and $\delta_c$ 
introduced in the slicing method.  In Fig.\ \ref{deltas}, we present the 
dependency of 
both the $2$-body and $3$-body pieces of the ${\cal O} (\alpha_s)$ 
contribution on $\delta_s$, keeping the ratio $\delta_s / \delta_c = 100$ 
fixed. These results are obtained for a specific choice of the ADD model 
parameters $M_D = 3$ TeV and $\delta = 4$. It can be seen from the 
figure that the sum of these two pieces is positive and is 
fairly stable for a wide range of $\delta_s$.  
The positive sum implies that the QCD corrections do 
enhance the leading order predictions.  In the rest of our work, 
we choose $\delta_s = 10^{-3}$ and $\delta_c = 10^{-5}$.
\begin{figure}[htb]
\centerline{
\epsfig{file=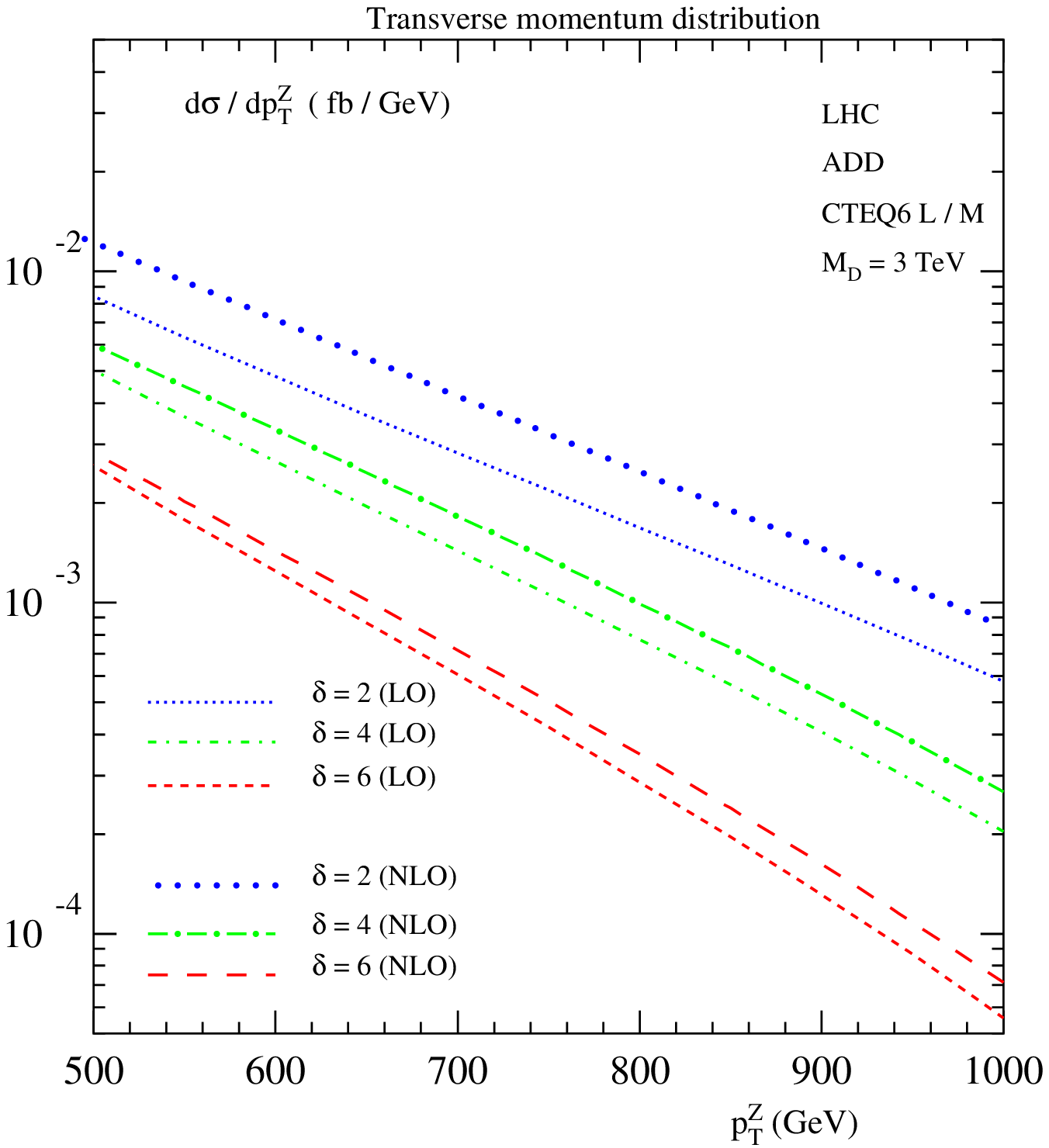,width=8cm,height=9cm,angle=0}
\epsfig{file=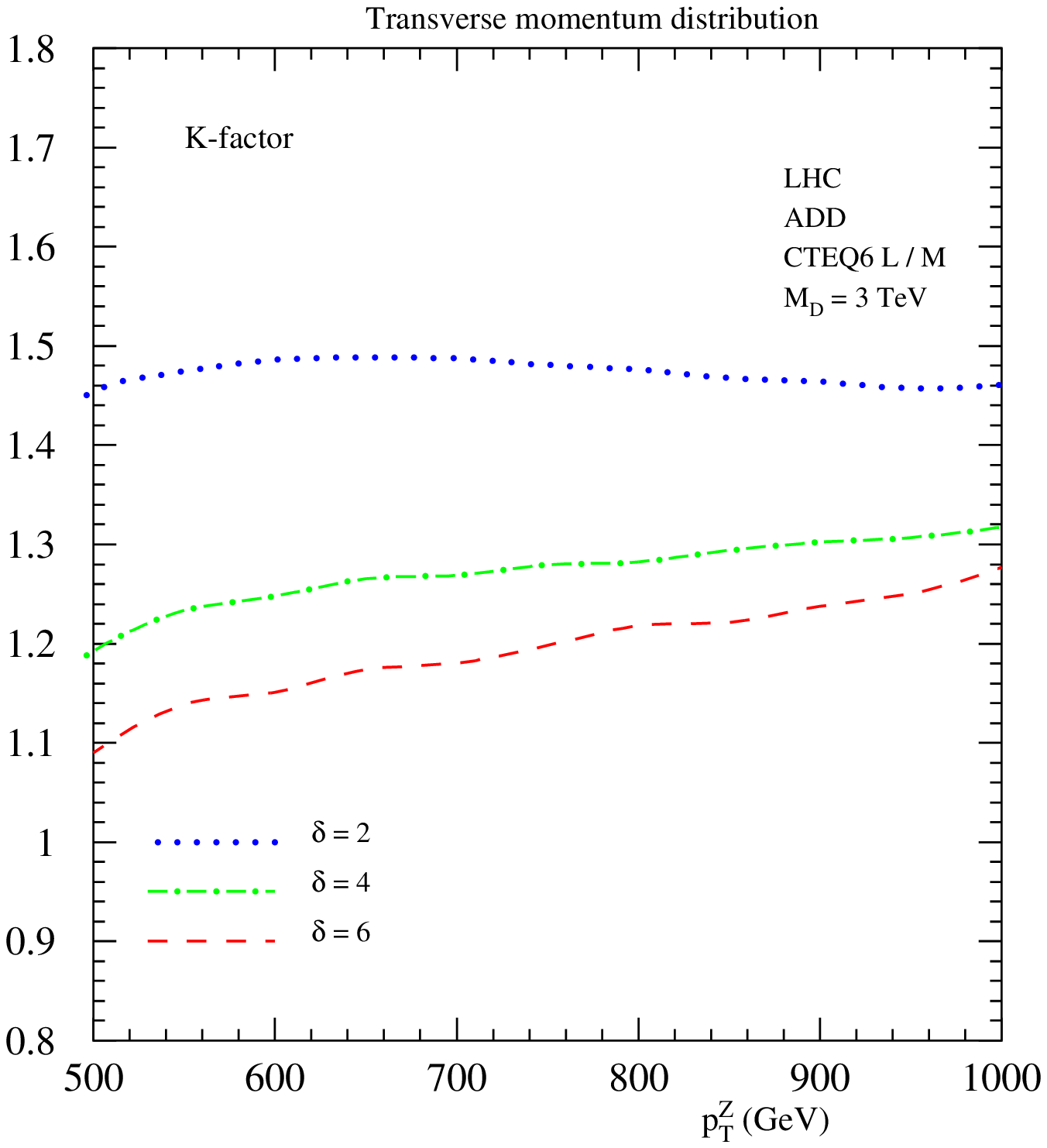,width=8cm,height=9cm,angle=0}}
\caption{Transverse momentum distribution of the $Z$-boson (left) and 
the corresponding K-factors (right) for the associated production of 
the $Z$-boson and the KK mode at the LHC, for $M_D = 3$ TeV and 
$\delta = 2, 4, 6$.}
\label{ptz}
\end{figure}

\begin{figure}[htb]
\centerline{
\epsfig{file=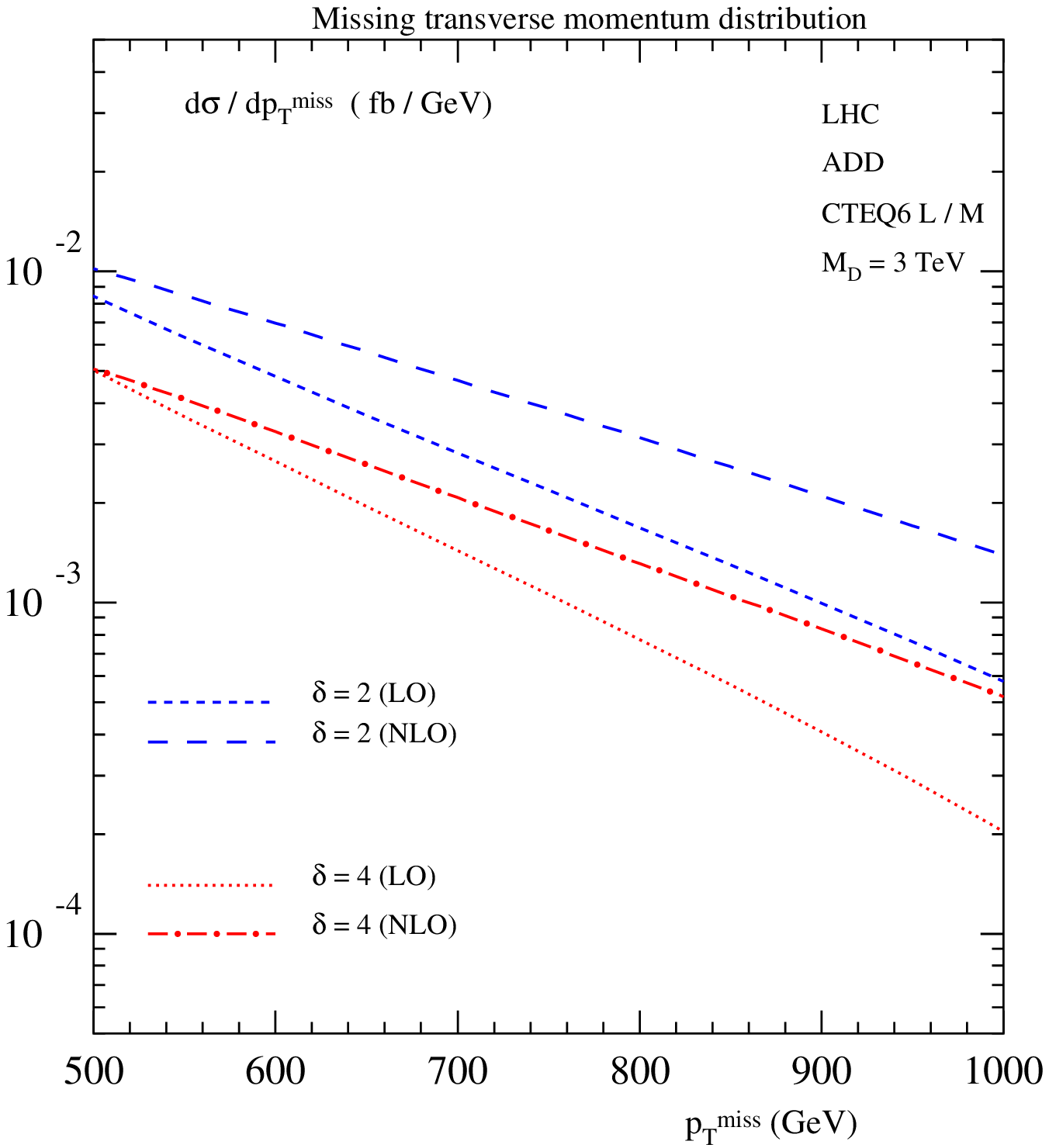,width=8cm,height=9cm,angle=0}
\epsfig{file=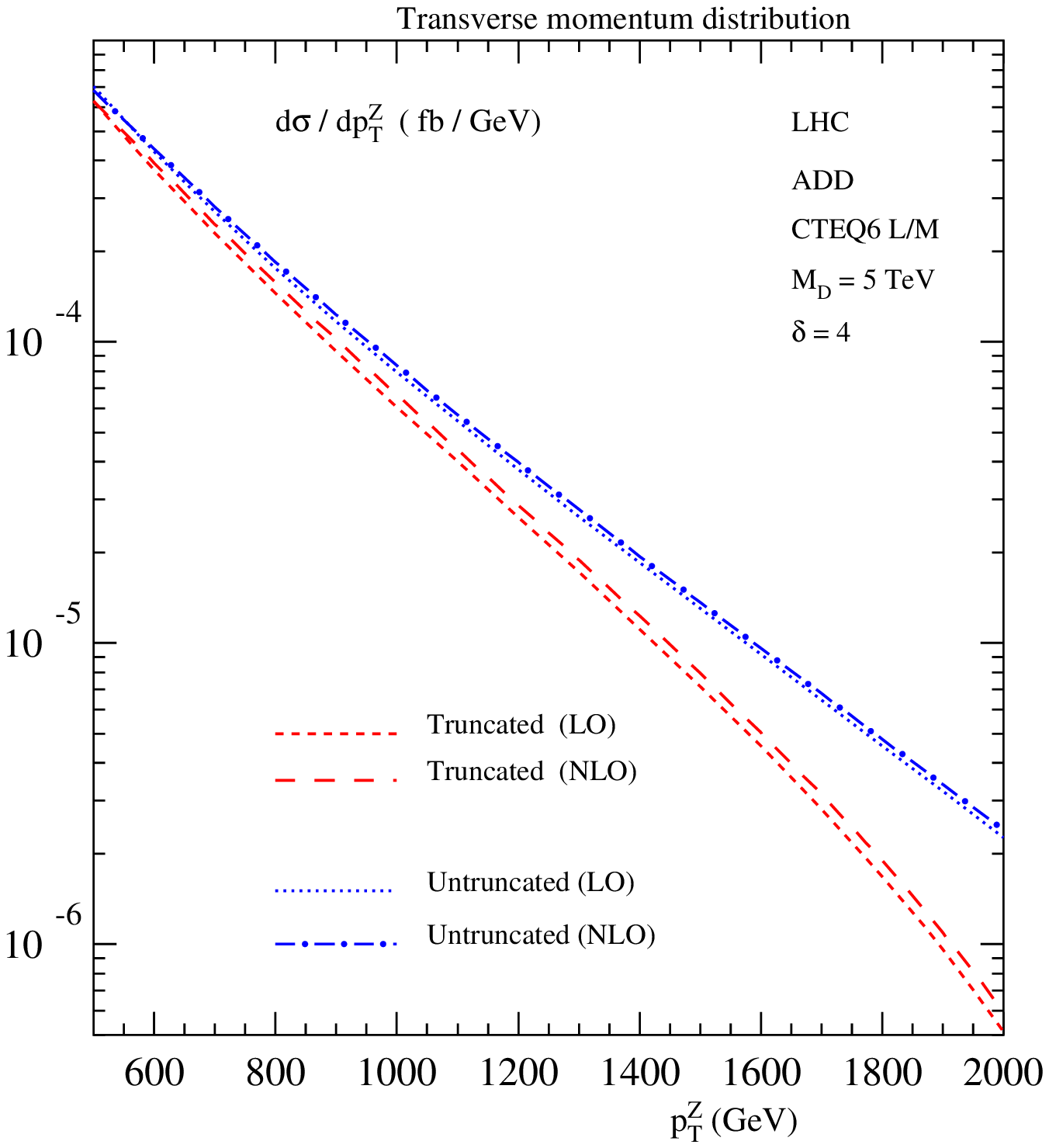,width=8cm,height=9cm,angle=0}}
\caption{Missing transverse momentum distribution (left) at the LHC for 
$M_D = 3$ and for $\delta = $ 2 and 4. 
The truncated as well as the un-truncated transverse momentum distributions
of the $Z$-boson (right) at the LHC for $M_D = 5$ and for $\delta = 4$ for 
both LO and NLO.}
\label{trunc}
\end{figure}

In Fig.\ \ref{ptz}, we present the transverse momentum distribution 
of the $Z$-boson at the LHC to NLO in QCD and its dependence on the 
number of extra dimensions $\delta$ for $M_D = 3$ TeV.  In each of the 
distributions corresponding to $\delta = 2, 4, 6$ the QCD effects are 
found to have increased the leading order predictions considerably.  In 
this transverse momentum distribution of the $Z$-boson, the $K$-factor, 
defined as the ratio of NLO cross sections to the LO ones, is found to 
increase with $p_T^Z$ and vary from 1.1 to 1.46 depending on the number 
of extra dimensions $\delta$.  In a similar way, the missing transverse 
momentum distribution is shown in the left panel of Fig.\ \ref{trunc}.

The ADD model which is an extension of the SM to address the hierarchy
problem is an effective theory and the UV completion of the TeV scale
gravity has to be quantified.  For the real KK mode production 
process, the kinematical constraint discussed earlier, provides a natural
UV cutoff on the integration over the n-sphere, but the hard scattering 
scales involved at the LHC energies can be close to the fundamental 
scale $M_D$.  To study the sensitivity of our results close to the UV 
region \cite{GRW}, we compare at LO and NLO the $p_T$ distribution of Z-boson 
wherein the invariant mass of the KK mode and Z-boson $Q_{ZG}$ is
(a) computed only when $Q_{ZG} < M_D$ ({\it truncation}) and
(b) computed for all possible values of $Q_{ZG}$ ({\it un-truncation}).
In the right panel of Fig.\ \ref{trunc}, we compare the results for
the ADD model parameters, $M_D=5$ TeV and $\delta=4$ at LO and NLO for
the $p_T$ distribution.  As compared to the un-truncated distribution, 
the percentage difference is tabulated below for LO and NLO for a 
few values of $p_T^Z$.  
\begin{center}
\begin{tabular}{ccc}
\hline
$p_T^Z$ (GeV) & LO   & NLO \\
\hline
500     & 11   & 7.8 \\
1000    & 23.5 & 20.2\\
1500    & 44.9 & 41.8\\
\hline
\end{tabular}
\end{center}
The difference between the un-truncated and truncated results 
become larger with (a) increase in $p_T^Z$, (b) increase in 
the number of extra dimensions $\delta$ and (c) decrease in the 
fundamental scale $M_D$.  Further it is seen that the NLO QCD 
corrections do decrease the difference between the 
un-truncated and truncated results.

\begin{figure}[htb]
\centerline{
\epsfig{file=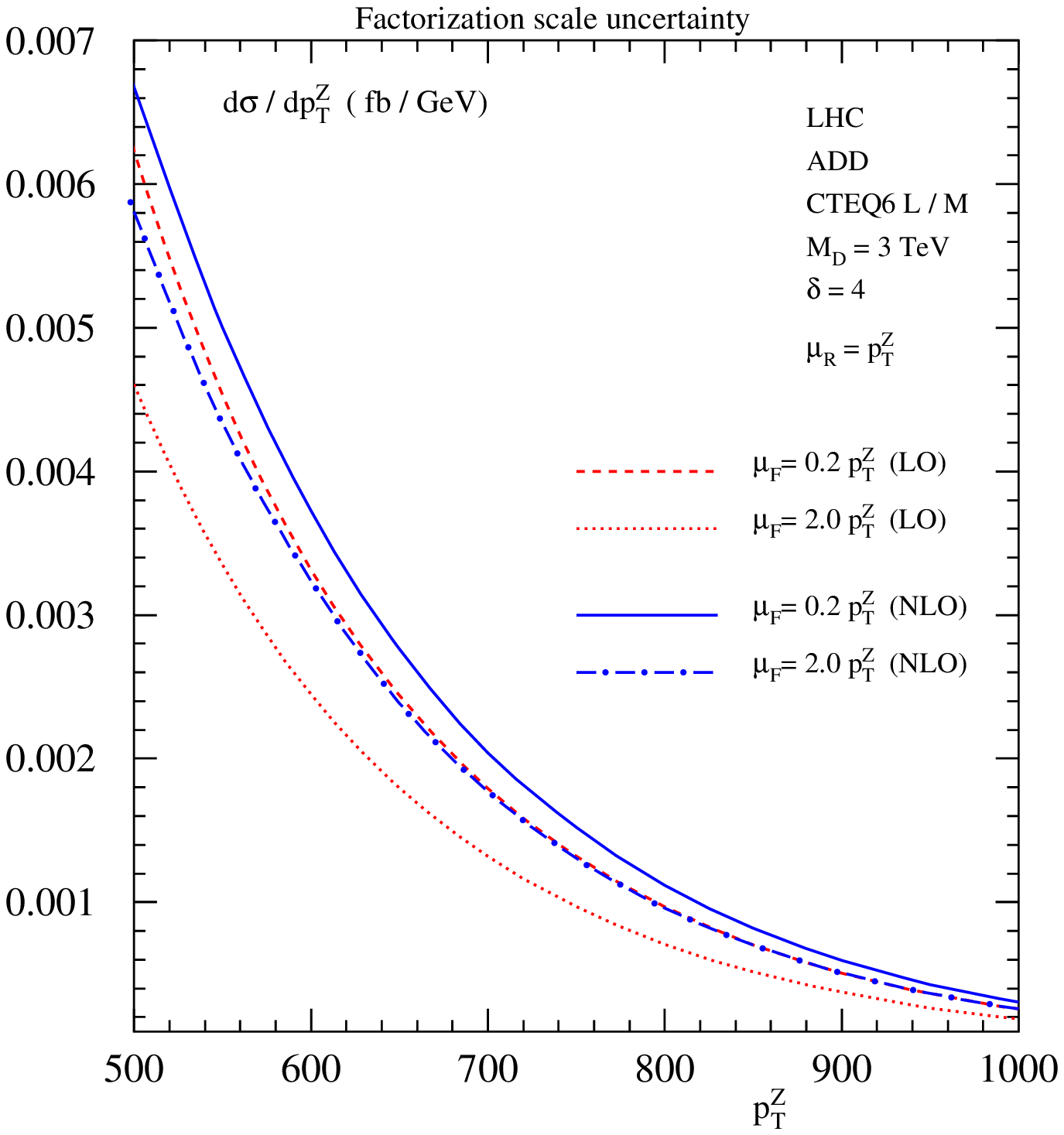,width=8cm,height=9cm,angle=0}
\epsfig{file=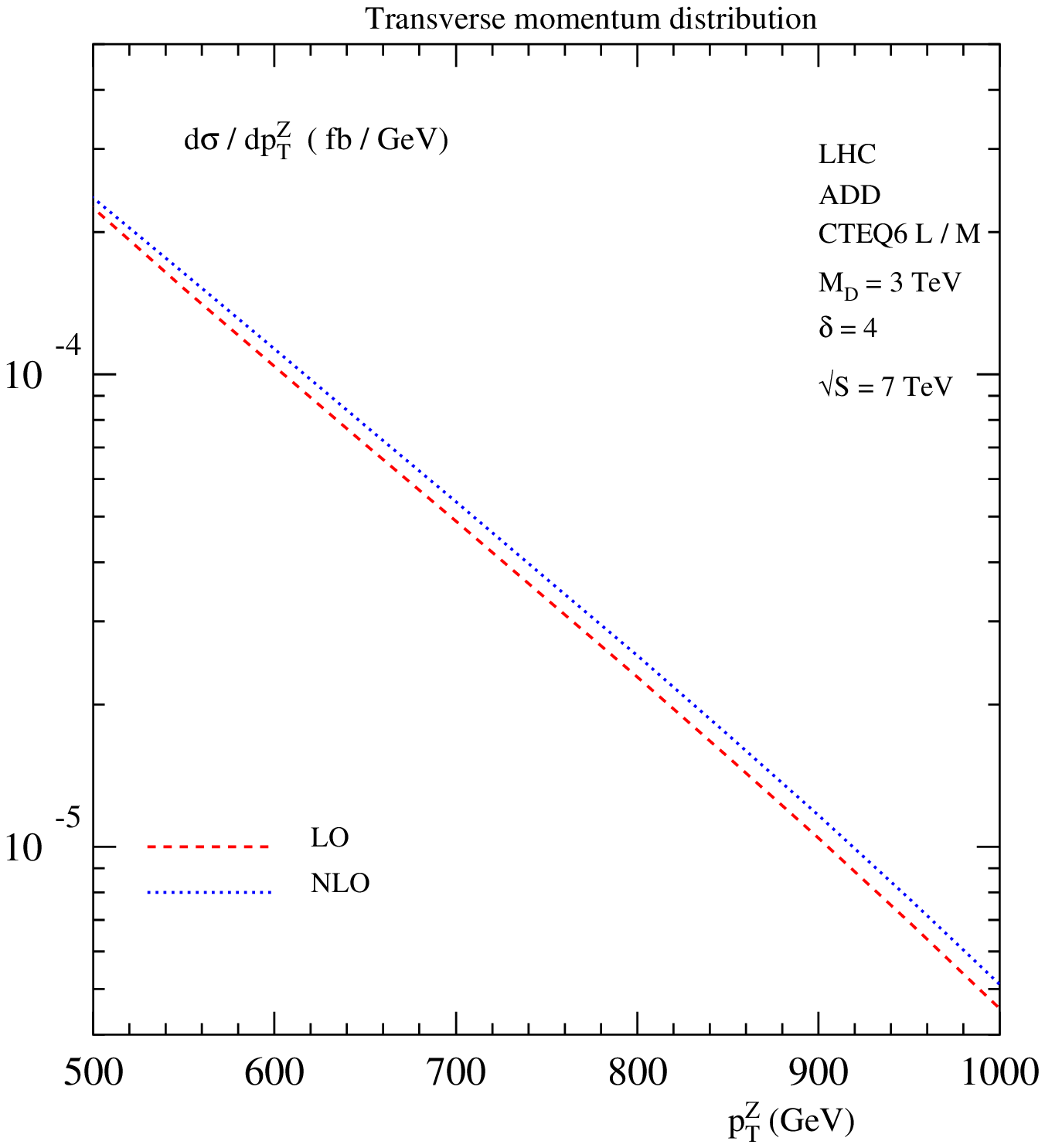,width=8cm,height=9cm,angle=0}}
\caption{Scale uncertainty in the transverse momentum 
distribution of the $Z$-boson at the LHC (left), for a variation of 
the factorization scale $\mu_F$ in the range $[0.2, 2.0]~p_T^Z$
and for the choice of $M_D = 3$ TeV and $\delta = 4$.
The transverse momentum distribution of the Z-boson
for $\sqrt{S}=7$ TeV at the LHC (right).
}
\label{scale}
\end{figure}

Finally, we study the dependence of both LO and NLO cross sections
on the factorization scale $\mu_F$ by varying it from $0.2~p_T^Z$ 
to $2.0~p_T^Z$.  One of the motivations for the computation of the 
QCD corrections is to minimise the scale uncertainties by computing 
the cross sections to higher orders in the perturbation theory.  As 
expected, the scale uncertainties in the leading order predictions 
are considerably decreased after incorporating the one-loop QCD 
corrections to the associated production of the $Z$-boson and the KK 
modes at the LHC.  The results are shown in the Fig.\ \ref{scale} 
(left) for the case of transverse momentum distribution of the 
$Z$-boson, for $M_D = 3$ TeV and $\delta = 4$.  In Fig.\ \ref{scale}
(right) we have also plotted the $p_T$ distribution for the current
LHC energies of $\sqrt{S} =7$ TeV.  The K-factor ranges from 1.05 to 
1.13 for $500 < p_T^Z < 1000$ GeV.  The contribution of $q (\bar q) ~g$ 
subprocess for the $\sqrt{S} =7$ TeV is much smaller than the contribution
at $\sqrt{S} =14$ TeV and that accounts for the much lower K-factor.  
\begin{figure}[htb]
\centerline{
\epsfig{file=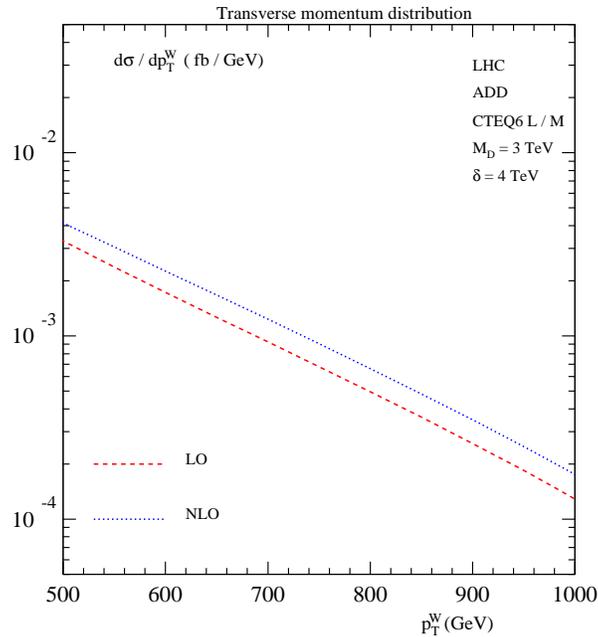,width=8cm,height=9cm,angle=0}}
\caption{ The $p_T$ distribution of $W^-$ in association 
with KK modes at the LHC with $\sqrt{S}=14$ TeV.
}
\label{W-}
\end{figure}

It is interesting to note here that a similar analysis goes through
for the associated production of $W^\pm$ and the KK modes in the 
large extra dimensional model at the LHC.  The difference lies both in 
the couplings of the quarks to the weak bosons and in the respective 
parton fluxes to be convoluted with the partonic cross sections.  
In Fig.\ \ref{W-} we have plotted the $W^-$ distribution for $\sqrt{S}=
14$ TeV at LO and NLO.   For the $p_T^W$ distribution the K-factor
is in the range 1.25 - 1.37 for $500 < p_T^W < 1000$ GeV.  A complete
analysis of the associated production of $W^\pm$ in association with
KK modes of the ADD model will be presented in the longer version
\cite{future}.

\section{Conclusion}
To conclude, we have computed the NLO QCD corrections to the associated 
production of the vector boson at the LHC, using semi-analytical two cutoff 
phase space slicing method.  
We have presented results for $\sqrt S =$ 14 and 7 TeV.
Our results are checked for the stability 
against the variation of the slicing parameters $\delta_s$ and $\delta_c$.  
We have studied the truncated as well as the un-truncated transverse 
momentum distributions of the $Z$-boson, together with the missing 
transverse momentum distribution and their dependence on the number 
of extra dimensions $\delta$.  The NLO QCD corrections are found to 
have not only enhanced the LO cross sections considerably,  with the 
K-factors ranging from 1.1 to 1.46 depending on the $\delta = 2, 4, 
6$ for $M_D = 3$ TeV, but also decreased their factorization scale 
uncertainties significantly.

\section*{Acknowledgments}
The work of V.R. and M.C.K. has been partially supported by funds made 
available to the Regional Centre for Accelerator based Particle Physics 
(RECAPP) by the Department of Atomic Energy, Govt. of India.  We would 
like to thank the cluster computing facility at Harish-Chandra Research 
Institute where part of computational work for this study was carried 
out.  S.S. would like to thank UGC, New Delhi for financial support. 
S.S. would also like to thank RECAPP center for his visit, where part of
the work was done.

\end{document}